\documentclass[a4paper,11pt]{article}

\usepackage{jcappub} 
\usepackage[T1]{fontenc} 
\usepackage{graphicx}
\usepackage{caption}
\usepackage{subcaption}
\usepackage{float}
\usepackage{slashed}

\title{\boldmath Constraints on long-lived electrically charged massive particles from anomalous strong lens systems}

\author[a]{Ayuki Kamada,}
\author[b]{Kaiki Taro Inoue,}
\author[c,d]{Kazunori Kohri,}
\author[e]{and Tomo Takahashi}

\affiliation[a]{Institute for Basic Science, Center for Theoretical Physics of the Universe, Daejeon 34051, South Korea}
\affiliation[b]{Faculty of Science and Engineering, Kindai University, Higashi-Osaka, Osaka, 577-8502, Japan}
\affiliation[c]{Institute of Particle and Nuclear Studies, KEK, 1-1 Oho, Tsukuba, Ibaraki 305-0801, Japan}
\affiliation[d]{The Graduate University for Advanced Studies (SOKENDAI), 1-1 Oho, Tsukuba, Ibaraki 305-0801, Japan}
\affiliation[e]{Department of Physics, Saga University, Saga 840-8502, Japan}

\emailAdd{akamada@ibs.re.kr}
\emailAdd{kinoue@phys.kindai.ac.jp}
\emailAdd{kohri@post.kek.jp}
\emailAdd{tomot@cc.saga-u.ac.jp}

\abstract{
We investigate anomalous strong lens systems, particularly the effects of weak lensing by structures in the line of sight, in models with long-lived electrically charged massive particles (CHAMPs).
In such models, matter density perturbations are suppressed through the acoustic damping and the flux ratio of lens systems are impacted, from which we can constrain the nature of CHAMPs. 
For this purpose, first we perform $N$-body simulations and develop a fitting formula to obtain non-linear matter power spectra in models where cold neutral dark matter and CHAMPs coexist in the early Universe. 
By using the observed anomalous quadruple lens samples, we obtained the constraints on the lifetime ($\tau_{\rm Ch}$) and the mass density fraction ($r_{\rm Ch}$) of CHAMPs. 
We show that, for $r_{\rm Ch}=1$, the lifetime is bounded as $\tau_{\rm Ch} < 0.96$\,yr (95\% confidence level), while a longer lifetime $\tau_{\rm Ch} = 10$\,yr is allowed when $r_{\rm Ch}  < 0.5$ at the 95\% confidence level.
Implications of our result for particle physics models are also discussed.
}

\newcommand{\f}{\frac}

\newcommand{\N}{\nonumber}
\newcommand{\BF}{\begin{figure}\begin{center}}
\newcommand{\EF}{\end{center}\end{figure}}
\newcommand{\BE}{\begin{equation}}
\newcommand{\EE}{\end{equation}}
\newcommand{\BEA}{\begin{eqnarray}}
\newcommand{\EEA}{\end{eqnarray}}

\newcommand{\tr}{\textrm}
\newcommand{\IG}{\includegraphics}

\begin{document}
\maketitle

\def\aj{AJ}%
\def\actaa{Acta Astron.}%
\def\araa{ARA\&A}%
\def\apj{ApJ}%
\def\apjl{ApJ}%
\def\apjs{ApJS}%
\def\ao{Appl.~Opt.}%
\def\apss{Ap\&SS}%
\def\aap{A\&A}%
\def\aapr{A\&A~Rev.}%
\def\aaps{A\&AS}%
\def\azh{AZh}%
\def\baas{BAAS}%
\def\bac{Bull. astr. Inst. Czechosl.}%
\def\caa{Chinese Astron. Astrophys.}%
\def\cjaa{Chinese J. Astron. Astrophys.}%
\def\icarus{Icarus}%
\def\jcap{J. Cosmology Astropart. Phys.}%
\def\jrasc{JRASC}%
\def\mnras{MNRAS}%
\def\memras{MmRAS}%
\def\na{New A}%
\def\nar{New A Rev.}%
\def\pasa{PASA}%
\def\pra{Phys.~Rev.~A}%
\def\prb{Phys.~Rev.~B}%
\def\prc{Phys.~Rev.~C}%
\def\prd{Phys.~Rev.~D}%
\def\pre{Phys.~Rev.~E}%
\def\prl{Phys.~Rev.~Lett.}%
\def\pasp{PASP}%
\def\pasj{PASJ}%
\def\qjras{QJRAS}%
\def\rmxaa{Rev. Mexicana Astron. Astrofis.}%
\def\skytel{S\&T}%
\def\solphys{Sol.~Phys.}%
\def\sovast{Soviet~Ast.}%
\def\ssr{Space~Sci.~Rev.}%
\def\zap{ZAp}%
\def\nat{Nature}%
\def\iaucirc{IAU~Circ.}%
\def\aplett{Astrophys.~Lett.}%
\def\apspr{Astrophys.~Space~Phys.~Res.}%
\def\bain{Bull.~Astron.~Inst.~Netherlands}%
\def\fcp{Fund.~Cosmic~Phys.}%
\def\gca{Geochim.~Cosmochim.~Acta}%
\def\grl{Geophys.~Res.~Lett.}%
\def\jcp{J.~Chem.~Phys.}%
\def\jgr{J.~Geophys.~Res.}%
\def\jqsrt{J.~Quant.~Spec.~Radiat.~Transf.}%
\def\memsai{Mem.~Soc.~Astron.~Italiana}%
\def\nphysa{Nucl.~Phys.~A}%
\def\physrep{Phys.~Rep.}%
\def\physscr{Phys.~Scr}%
\def\planss{Planet.~Space~Sci.}%
\def\procspie{Proc.~SPIE}%
          
\flushbottom

\section{Introduction}
\label{sec:intro}
Long-lived electrically charged massive particles (CHAMPs)\,\cite{Sigurdson:2003vy, Profumo:2004qt, Kohri:2009mi, Kamada:2013sh} are suggested to resolve the failure of the standard $\Lambda$CDM ($\Lambda$: dark energy; CDM: cold dark matter) model in reproducing the observed number of satellite galaxies in the Milky Way\,\cite{Moore:1999nt, Kravtsov:2009gi} with satisfying cosmological bounds, e.g., from big-bang nucleosynthesis\,\cite{Pospelov:2006sc, Steffen:2006wx, Hamaguchi:2007mp, Bird:2007ge, Kawasaki:2007xb, Kawasaki:2008qe}, cosmic microwave background (CMB) anisotropies\,\cite{Kamada:2016qjo}, and magnetic fields in galaxy clusters\,\cite{Kadota:2016tqq} (for other bounds, see \cite{Kamada:2016qjo} and references therein).  
The Coulomb interaction ties CHAMPs to baryons until CHAMPs decay into neutral, stable, and massive particles, which account for part of the present DM mass density.  
CHAMPs are prevented from falling into the bottom of the primordial gravitational potential by the induced pressure.  
The resultant matter power spectra are suppressed below the comoving horizon length scale at the CHAMP decay when compared to that in the standard CDM model.

As a concrete example of long-lived CHAMPs in particle physics models, especially when we consider supersymmetric (SUSY) models, it is known that a scalar partner of charged leptons tends to have a electroweak-scale mass and a long lifetime. 
Two possible cases have been known, where the scalar charged leptons can indeed become long-lived in the minimal-supersymmetric standard model (MSSM) (see, e.g., \cite{Feng:2003xh, Feng:2003uy, Feng:2004zu, Feng:2004mt}). 
First, the lightest SUSY particle (LSP) can be the gravitino with the next-to-LSP (NLSP) being a scalar charged lepton, e.g., the scalar tau lepton (i.e., stau) in the gauge-mediated SUSY breaking models\,\cite{Dine:1994vc, Dine:1995ag}. 
The NLSP stau decays into the gravitino LSP only through the gravitational interaction and thus the electroweak-scale stau has a long lifetime. 
Second, in the case of a degenerate mass spectrum between the stau ($\tilde{\tau}$) NLSP and the neutralino ($\chi$) LSP, the stau can be long-lived kinematically due to the smallness of the mass difference. 
In the latter model, it is remarkable that the lifetime can be of the order of $0.1 \text{--} 10$\,yr with their small mass difference being $\Delta m < 50$\,MeV and their large absolute masses being $\gtrsim 100$\,GeV\,\cite{Jittoh:2007fr}. 
In this case, a possible decay mode should be only into $\tilde{\tau} \to \chi + e + \nu_e + \nu_{\tau}$ with a negligible kinetic energy of the daughter electron ($e$)\,\cite{Jittoh:2007fr}.  
Therefore, the decay of the stau NLSP does not induce any significant photodissociations of background $^4$He or D\,\cite{Kawasaki:2004qu, Jedamzik:2006xz} at a low temperature of $T \lesssim 10$\,keV, which agrees with the observed light element abundances.

Even if we consider a more general case that is no longer based on the MSSM, a degenerate mass spectrum between the CHAMP and its daughter particle kinematically induces a long lifetime of the CHAMP, while neither high-energy charged leptons nor photons are emitted through the decay by a small mass difference.
Remarkably, the lifetime of the CHAMP tends to be longer for a larger mass of the CHAMP with the mass difference being kept fixed. 
For example, in order to obtain a conservative bound on the CHAMP, it would be possible to consider a mass spectrum in which $\Delta m$ is smaller than ${\cal O}(10)$\,MeV by which any photodissociations of the light elements are not induced, and the free-streaming length of the daughter particle is too short to erase the density perturbations below $k = {\mathcal O}(10^{4}) \,h/$Mpc. 
In this paper therefore, we consider such conservative setups for the decaying CHAMP models by assuming that negligible amounts of energetic electromagnetic particles are emitted with a negligible kinetic energy of the daughter particle.

We use the analyzing method in \cite{Inoue:2014jka, Kamada:2016vsc} though we consider long-lived CHAMP models instead of warm (WDM) or mixed (MDM) dark matter models.
In both the models the formation of subgalactic-scale ($k = {\mathcal O}(100)\, h$/Mpc) objects is suppressed.
One may wonder if we can map a derived constraint on MDM models, more specifically, the WDM mass and the mass density fraction of the warm component ($m_{\rm WDM}$ and $r_{\rm warm}$) to the lifetime and the mass density fraction of CHAMPs ($\tau_{\rm Ch}$ and $r_{\rm Ch}$) with a help of the relation derived in WDM models\,\cite{Kamada:2013sh}: $\tau_{\rm Ch} \simeq 1.5 \, {\rm yr} \, (1 \, {\rm keV} / m_{\rm WDM})^{8/3}$.
Our result actually confirms that this relation is valid between pure long-lived CHAMP models ($r_{\rm Ch} = 1$) and WDM models ($r_{\rm warm} = 1$).
Nevertheless, as we will see later, this mapping is not applicable between mixed long-lived CHAMP ($r_{\rm Ch} < 1$) models and MDM ($r_{\rm warm} < 1$) models.
This is virtually because of the difference in suppression mechanisms between long-lived CHAMP models and MDM models: 
the acoustic damping through the (indirect) interaction with relativistic species and the collisionless damping through the free-streaming of the warm component, respectively.
Thus we perform independent analysis in long-lived CHAMP models in this paper, even though the methodology itself is close to that in \cite{Inoue:2014jka, Kamada:2016vsc}.

In order to constrain long-lived CHAMP models, we use quadruply lensed quasar-galaxy systems that show anomalous flux ratios in lensed images; the flux ratios of lensed images disagree with the prediction of the best-fit lens models with a smooth potential whose variation length scale is larger than the separation between the lensed images. 
It has been argued that such anomalies are caused by subhalos in lens galaxies\,\cite{Mao:1997ek, Chiba:2001wk, Dalal:2001fq, Metcalf:2001es, Kochanek:2003zc, Metcalf:2003sz, Chiba:2005et, Sugai:2007ic, More:2008qm, Minezaki:2009ek, Xu:2009ch, Xu:2010gs, Inoue:2017zxb}.  
However, recent analyses showed that the contribution from intervening halos are significant\,\cite{Metcalf:2004eh, Xu:2011ru, Inoue:2012px, Takahashi:2013sna}, while the predicted subhalo population is too low to explain the observed anomalous flux ratios\,\cite{Maccio:2005bj, Amara:2004dr, Xu:2009ch, Xu:2010gs, Chen:2008vt, Chen:2011wc}. 
The contribution from subhalos is turned out to be just $\sim 30\%$ of the total that includes contribution from intervening structures such as halos, voids, and filaments along the line of sight\,\cite{Inoue:2014mla, Inoue:2016mqz}. 
In fact, a sign of perturbations by locally negative density perturbations has been observed in the quadruply lensed submillimeter galaxy (SDP.81)\,\cite{Inoue:2015lma}. 
In our analysis, we assume that the observed anomalous flux ratios are mainly caused by intervening structures rather than subhalos in lens galaxies\,\cite{Miranda:2007rb, Inoue:2012px}.

We use optical/near-infrared data for the positions of lensed images and the centroid of the foreground elliptical galaxy and mid-infrared/radio (cm) data for the flux of lensed images. 
Since the angular size of lensed images in the mid-infrared/radio bands is sufficiently larger than the Einstein angular radius of stars in the foreground galaxy, the change in the flux of lensed images due to microlenses is negligible in the mid-infrared/radio bands. 
We assume that the primary lens in each system is described by a smooth potential with an elliptical symmetry in the surface mass density projected onto the lens plane. 
We fit the data with a model that includes a primary lens with an external shear and possibly a companion galaxy if necessary. 
Then the magnification perturbation in each system is estimated from the residual in the flux of lensed images obtained from the fit. 
If the acoustic damping of small-scale density perturbations is too strong, then the magnification perturbation cannot be explained by structures in the line of sight. 
Thus we can constrain the lifetime and the mass density fraction of CHAMPs.

The organization of this paper is as follows.
In section\,\ref{sec:matpower} we study the behavior of density perturbations in both the linear and non-linear regimes.
There we describe the physical mechanism that results in the suppression of the linear matter power spectrum in long-lived CHAMP models.
A fitting formula of non-linear matter power spectra is derived.
By using the fitting formula we evaluate the second moment of the probability density function (PDF) of the magnification perturbation in section\,\ref{sec:lenanares}.
Then we derive the probability of reproducing the observed flux ratios in a given long-lived CHAMP model, which is used to constrain the long-lived CHAMP model parameters.
Section\,\ref{sec:concl} is devoted to concluding remarks. 
Throughout this paper, we take the cosmological parameters obtained from the observed CMB anisotropies (``{\it Planck} + WP'' (WP: WMAP polarization)\,\cite{Ade:2013zuv}) to be consistent with \cite{Inoue:2014jka, Kamada:2016vsc}:
the matter mass density at present, $\Omega_{m, 0}=0.3134$; the baryon mass density, $\Omega_{b, 0}=0.0487$; the cosmological constant, $\Omega_{\Lambda, 0}=0.6866$; the Hubble constant, $H_0=67.3$ $(=100h)\, {\rm km / s / Mpc}$; the spectral index, $n_{s}=0.9603$; the root-mean-square (rms) amplitude of matter density perturbations at $8 h^{-1}\, {\rm Mpc}$, $\sigma_8=0.8421$.

\section{Matter power spectra}
\label{sec:matpower}

\subsection{Linear matter power spectra}
\label{subsec:linpower}
We suitably modify the public code \texttt{CAMB}\,\cite{Lewis:1999bs} to incorporate the evolution equations of the long-lived CHAMP perturbations given in \cite{Sigurdson:2003vy, Profumo:2004qt}.
Here the tight coupling between CHAMPs and baryons is assumed, which is valid as long as $m_{\rm Ch} < 10^{8}$\,GeV\,\cite{Kamada:2016qjo}.
In this mass range, the matter power spectra in long-lived CHAMP models are parametrized by the lifetime and the mass density fraction: 
$\tau_{\rm Ch}$ and $r_{\rm Ch} = \Omega_{{\rm Ch}, 0} / \Omega_{c, 0}$, where $\Omega_{{\rm Ch}, 0}$ is the CHAMP mass density evaluated assuming that it is stable until today even for the case with the lifetime ($\tau_{\rm Ch}$) is shorter than the age of the Universe. 
$\Omega_{c, 0}$ is the CDM mass density at present.
We assume that some CDM-like particles account for the rest of the observed DM mass density: $1 - r_{\rm Ch}$.
Note that this CDM-like particle could be identical to the neutral decay product of the CHAMP, but their origins and thermal histories are completely different; the CDM-like particle should exist as a stable and cold component of the Universe at least around and after the horizon entry of the perturbations of interest.
A cutoff scale can be estimated at the comoving horizon scale at the CHAMP decay:
\BE
\label{eq:cutoffCh}
k_{\rm Ch} = a H |_{t = \tau_{\rm Ch}} \simeq 0.18 \, {\rm / Mpc} \left( \frac{1 \, {\rm yr}}{\tau_{\rm Ch}} \right)^{1/2} \,,
\EE
with the scale factor $a$ being normalized such that $a = 1$ at present and the Hubble expansion rate being $H$.

In general $r_{\rm Ch}$ can be larger than unity with the present mass density of the neutral decay product equal to or smaller than the observed DM mass density: $\Omega_{{\rm daughter}, 0} / \Omega_{c, 0} \leq 1$.  
This is because the CHAMP mass can be much larger than the mass of the neutral CHAMP decay product.  
In this paper, on the other hand, we focus on the case that the two masses are degenerate: $m_{\rm Ch} \simeq m_{\rm daughter}$.
This choice has two advantages: natural explanation of the CHAMP long lifetime and harmlessness of charged decay product.  
One may wonder how we can realize a relatively long-lived massive particle.  
One possible reason is that the electroweak-scale CHAMP decays into the neutral decay product through a superweak interaction, say, gravitationally (see \cite{Feng:2003xh, Feng:2003uy, Feng:2004zu, Feng:2004mt} for supersymmetric realizations).  
When the CHAMP mass is much larger than the electroweak scale, even a gravitational decay is not sufficient to make the CHAMP stable over ${\mathcal O}(1)$\,yr.  
The small mass difference between the CHAMP and the neutral decay product restricts the final-state phase space and lengthens the CHAMP lifetime.  
We also care about the energy injection from the CHAMP decay into the standard model plasma, which may be disfavored since they could destroy the light element produced from the big-bang nucleosynthesis\,\cite{Kawasaki:2004qu, Jedamzik:2006xz, Kawasaki:2008qe}, and/or distort the CMB black body spectrum to generate the $y$- and $\mu$-distortions\,\cite{Hu:1993gc, Chluba:2011hw}.

Figure\,\ref{fig:linpower} shows the linear matter power spectra extrapolated to the present time ($z=0$) in some long-lived CHAMP, WDM, and MDM models.
For notational simplicity, we denote a long-lived CHAMP model with the lifetime of $\tau_{\rm Ch}$ and the mass density fraction of $r_{\rm Ch}$ as CH($\tau_{\rm Ch}$, \,$r_{\rm Ch}$), where $\tau_{\rm Ch}$ is measured in units of year as shown in Table\,\ref{tab:modpara}.
For reference we also show the linear matter power spectra in MDM models, where the warm component follows the Fermi--Dirac distribution with the spin being one half and is characterized by the warm particle mass and the mass density fraction: $m_{\rm WDM}$ and $r_{\rm warm} = \Omega_{{\rm warm},0} / \Omega_{c, 0}$.
A cutoff scale in WDM models is given by the Jeans scale at matter-radiation equality:
\BE
\label{eq:cutoffJ}
k_{\rm J} = a \sqrt{\f{4 \pi G \rho_{\rm M}}{\sigma^2}} \bigg|_{t=t_{\rm eq}}  = 20/{\rm Mpc} \, \left( \f{m_{\rm WDM}}{1\,{\rm keV}} \right)^{4/3} \,,
\EE
with the velocity dispersion being $\sigma^2$\,\cite{Kamada:2013sh}.
The face values of eqs.\,(\ref{eq:cutoffCh}) and (\ref{eq:cutoffJ}) do not completely coincide with the cutoff scales read from calculated linear matter power spectra.
Nevertheless the scaling with the parameter is found appropriate.
It is suggested that we use $k_{\rm cut} \simeq 12 k_{\rm Ch} \simeq k_{\rm J} / 12$ as an empirical cutoff scale\,\cite{Kamada:2013sh}.
The mapping of the model parameters between pure long-lived CHAMP models and WDM models can be done with the following relation:
\BE
\label{eq:linrelation}
\tau_{\rm Ch} \simeq 1.5 \, {\rm yr} \, \left( \frac{1 \, {\rm keV}}{m_{\rm WDM}} \right)^{8/3} \,.
\EE

As seen from figure\,\ref{fig:linpower}, the CH(1,\,1) model (orange) shows an oscillation, which is not seen in the WDM model with $m_{\rm WDM} = 1.3$\, keV (blue); the density perturbation crosses zero around $k \simeq 14\,h$/Mpc and the peak at $k \simeq 20\,h$/Mpc has the opposite sign to the perturbations at $k < 10\,h$/Mpc.
The oscillation is an imprint of the acoustic oscillation between CHAMPs and baryons that takes place until the CHAMP decay.
 
\begin{table}
\centering
\begin{tabular}{lcc}
\hline \hline
Model & $\tau_{\rm Ch}$ [yr] & $r_{\rm Ch}$ \\
\hline
CDM & - & 0 \\
\hline
CH(0.1,\,0.1) & 0.1 & 0.1 \\
CH(1,\,0.1) & 1 & 0.1 \\
CH(10,\,0.1) & 10 & 0.1 \\
CH(0.1,\,0.5) & 0.1 & 0.5 \\
CH(1,\,0.5) & 1 & 0.5 \\
CH(10,\,0.5) & 10 & 0.5 \\
CH(0.1,\,0.7) & 0.1 & 0.7 \\
CH(1,\,0.7) & 1 & 0.7 \\
CH(10,\,0.7) & 10 & 0.7 \\
CH(0.1,\,0.85) & 0.1 & 0.85 \\
CH(1,\,0.85) & 1 & 0.85 \\
CH(10,\,0.85) & 10 & 0.85 \\
CH(0.1,\,1) & 0.1 & 1 \\   
CH(1,\,1) & 1 & 1 \\  
CH(10,\,1) & 10 & 1 \\   
\hline
\end{tabular}
\caption{\label{tab:modpara} Simulated models.}
\end{table}

Let us highlight another important feature by comparing the CH(10,\,0.5) model (green) and the MDM model with $(m_{\rm WDM} \, [{\rm keV}], r_{\rm warm}) = (0.5, 0.5)$ (red).
The suppression at smaller length scales is weaker in the long-lived CHAMP models than in the MDM models even with the mass density fraction being identical: $r_{\rm Ch} = r_{\rm warm}$.
This is because the suppression mechanisms are different between them.
Long-lived CHAMPs suppress the density perturbations of the neutral decay product through the acoustic damping discussed above, while in MDM models a collisionless damping is essential and the free-streaming of the warm component transfers the power of the density perturbation into the higher order terms of the Boltzmann hierarchy.
The former works in the radiation dominated era, or precisely speaking, until {\it baryon}s including CHAMPs dominate the entropy density or the long-lived CHAMP decays, while the latter works both in the radiation dominated era and in the matter dominated era. 
In the radiation dominated era the gravitational coupling between long-lived CHAMPs and the cold component is negligible since the gravitational potential is determined by the radiation energy density.
On the other hand, in MDM models, the gravitational force between the warm and cold components becomes important in the matter dominated era;
the velocity dispersion of the warm component prevents not only the warm component but also the cold one from growing gravitationally.
This observation is supported by figure\,\ref{fig:linpower}, which shows that the matter powers in the CH(10,\,0.5) model is a quarter $(= r_{\rm Ch}^{2})$ smaller than the CDM (black) at smaller length scales.

\BF
\IG[width=0.75\linewidth]{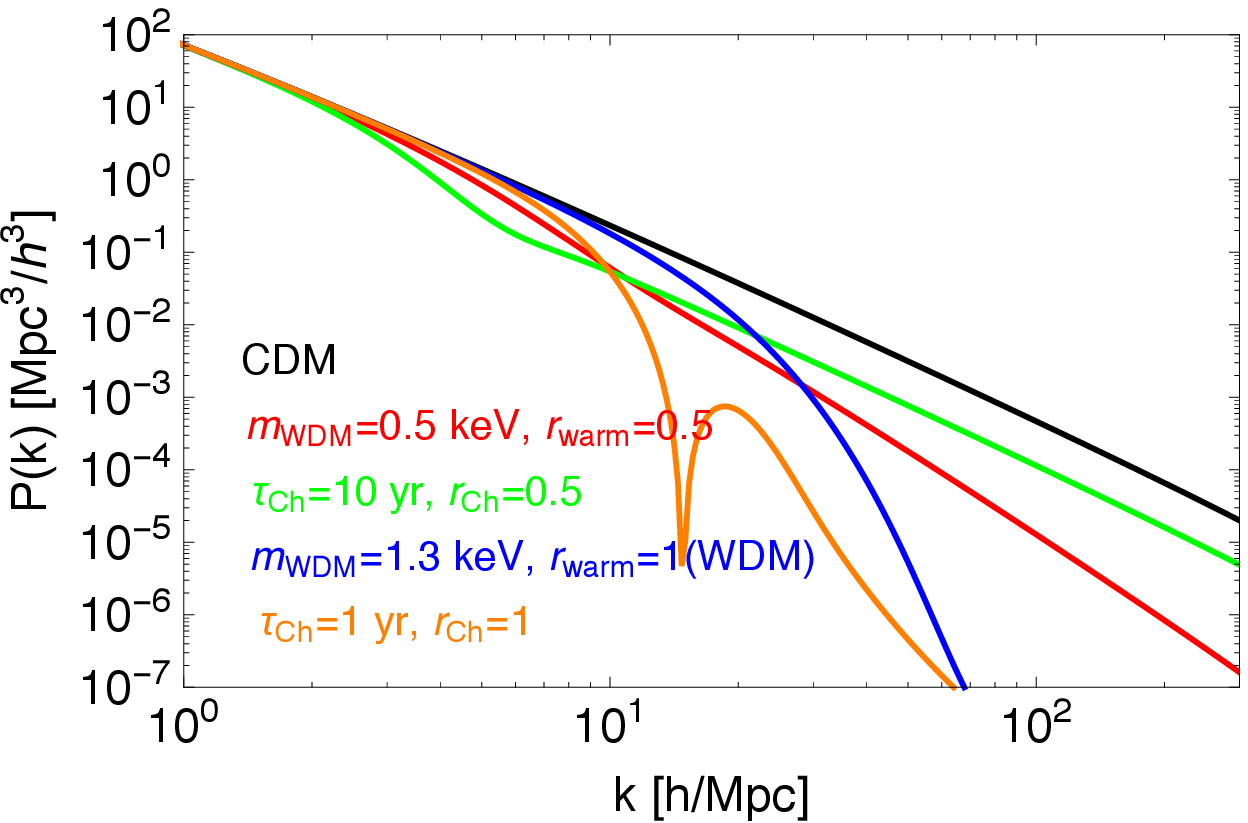}
\caption{Linear matter power spectra at $z=0$.
Among the models in table\,\ref{tab:modpara}, we compare models of CDM (black), CH(10,\,0.5) (green), and CH(1,\,1) (orange).
For reference we also show the MDM models with $(m_{\rm WDM} \, [{\rm keV}], r_{\rm warm}) = (0.5, 0.5)$ (red) and $(1.3, 1)$ (blue), which were respectively excluded at the 90 and 95\% confidence levels\,\cite{Inoue:2014jka, Kamada:2016vsc} by using the four anomalous samples of quadruple lenses that are also adopted in this paper.
}
\label{fig:linpower}
\EF

\subsection{Non-linear matter power spectra}
\label{subsec:nonlinpower}
We perform $N$-body simulations by using linear matter power spectra calculated in the models listed in table\,\ref{tab:modpara} to generate the initial conditions.
The public code \texttt{Gadget-2}\,\cite{Springel:2005mi} is used with the initial redshift of $z=49$ and the other setups summarized in table\,\ref{tab:simset}.
The simulation particles represent the whole matter distribution and are random samplings of its initial phase space distribution through the Zel'dovich approximation.
The realization of the random variable is identical among the simulated long-lived CHAMP models as well as the simulated MDM models investigated in \cite{Kamada:2016vsc}.

\begin{table}
\centering
\begin{tabular}{lccc}
\hline \hline
Setup & $L\,[{\rm Mpc}/h]$ & $N$ & $\epsilon\,[{\rm kpc}/h]$ \\
\hline
L5 & 5 & $512^{3}$ & 0.5 \\
\hline
L10 & 10 & $512^3$ & 1.0 \\
\hline 
HL10 & 10 & $1024^3$ & 0.5 \\
\hline
\end{tabular}
\caption{\label{tab:simset} Simulation setups. 
$L$ is the length on a side of the simulation box, $N$ is the number of the simulation particles, and $\epsilon$ is the gravitational softening length. 
HL10 is available only for the CDM, CH(1,\,0.85), CH(10,\,0.5), CH(0.1,\,1), and CH(1,\,1) models.
}
\end{table}

The direct use of the simulated matter distributions is preferable to calculate the magnification perturbations in long-lived CHAMP models.
The sampling given in table\,\ref{tab:modpara}, on the other hand, is not sufficiently dense to obtain a smooth constraint map in the $\tau_{\rm Ch}$-$r_{\rm Ch}$ plane.
Relying on the weak lens approximation, we can evaluate the magnification perturbations from the non-linear matter power spectra.
To this end we construct the fitting formula of the non-linear matter power spectra as a function of $\tau_{\rm Ch}$ and $r_{\rm Ch}$ through the following steps: 
we construct the fitting formula of the L5 and L10 simulated transfer functions ($T^{2} = P_{\rm Ch}/P_{\rm CDM}$); 
we obtain the fitting formula of the non-linear matter power spectra by multiplying the \texttt{halofit}\,\cite{Smith:2002dz} and $T^{2}$; 
and then we check the derived fitting formula with the HL10 simulations.

First we measure the power spectra of the simulated whole matter distributions in the L5 and L10 simulations in the pure long-lived CHAMP models ($r_{\rm Ch} = 1$) at $z=0,0.3,0.6,1,2$, and $3$.
Figure\,\ref{fig:transfc} shows the simulated transfer functions in the CH(0.1,\,1) and CH(1,\,1) models: $P_{\rm Ch}/P_{\rm CDM}|_{\rm L5, L10}$.
The L10 simulations show slightly larger matter powers than the L5 simulations.
This is because a smaller box size simulation generically misses effects of non-linear coupling between smaller and larger length-scale perturbations, which are more important in suppressed linear matter power models than in the standard CDM model.
On the other hand, the agreement between the L5 and L10 simulations is up to 10\% in the amplitude of the transfer function above the Nyquist wavenumber of the L5 simulation ($k_{\rm Nyq}=322 \, h/$Mpc) at $z < 2$ and sufficient for our purpose.

Interestingly the following fitting formula motivated by analogy to that in WDM models\,\cite{Inoue:2014jka} can reproduce the pure long-lived CHAMP simulations up to 10\% in the amplitude of the transfer function as shown in figure\,\ref{fig:transfc}:
\BE
\label{eq:transfc}
T^{2} (k_{\rm d}) = \frac{1}{(1 + k / k_{\rm d})^{0.7441}} \,, \quad
k_{\rm d} (\tau_{\rm Ch}) = 503.2 \, h / {\rm Mpc} \, \left( \frac{1 \, {\rm yr}}{\tau_{\rm Ch}} \right)^{0.9273} \,.
\EE
Here we vary and determine the overall factor and the scaling index of $k_{\rm d}$, which is a function of $\tau_{\rm Ch}$, by minimizing the residual:
\BE
\label{eq:res}
\sum_{\substack{r_{\rm Ch}=1 \\ {\rm models}}} \sum_{z \, {\rm bins}} \sum_{k \, {\rm bins}} \left( T^{2} - P_{\rm Ch}/P_{\rm CDM} |_{\rm L5} \right)^2 \,,
\EE
where $z \, {\rm bins}$, $z \in \{0,0.3,0.6,1,2,3\}$, and $k \, {\rm bins}$, $\log(k \, [h/{\rm Mpc}]) \in \{2.117,2.137,2.157,\cdots,2.477\}$ (total 20 bins), have a large number of samples and are not affected by cosmic variance.
Note that we can evaluate the overall factor and scaling index as $543.8$ and $0.8277$, respectively, by using the relation given in eq.\,(\ref{eq:linrelation}) though it is read from the linear matter power spectra.
These numbers are quite close to those obtained by the above fitting procedure, which support the suggestion of \cite{Kamada:2013sh}: the relation given in eq.\,(\ref{eq:linrelation}) map WDM models to pure long-lived CHAMP models even after the non-linear growth.

\BF
\IG[width=0.5\linewidth]{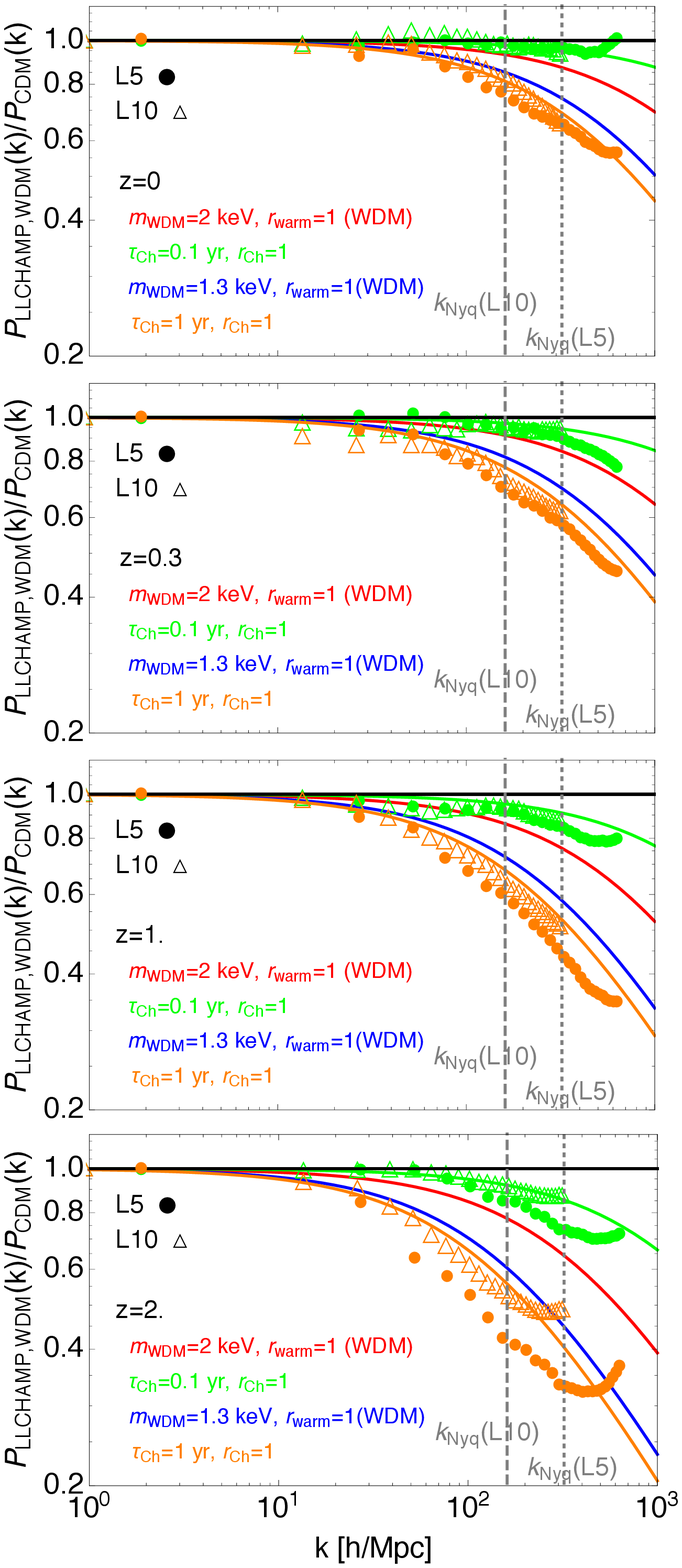}
\caption{Comparison of $T^{2}$ (see eq.\,(\ref{eq:transfc})) and $P_{\rm MDM}/P_{\rm CDM} |_{\rm L5, L10}$ at $z=0,0.3,1,$ and $2$. 
$T^{2}$ gives a reasonable fit up to 10\% in amplitude in the CH(0.1,\,1) (green) and CH(1,\,1) (orange) models at low redshifts ($z < 2$).
The fitting functions for the WDM models with $m_{\rm WDM} = 1.3$\,keV (blue) and $2$\,keV (red) (see \cite{Inoue:2014jka}) are also shown for reference.
We present the Nyquist wavenumbers of the L5 and L10 simulations (dashed lines), below which the measured matter power spectra are reliable. 
The upturns may originate from the discreteness effect of simulations, which is commonly seen in suppressed matter power models\,\cite{Wang:2007he}.
}
\label{fig:transfc}
\EF

Next we extend the above fitting formula to the transfer functions in mixed long-lived CHAMP models.
We have found that the following extension by analogy to that from WDM models to MDM models\,\cite{Kamada:2016vsc} provides a good fit at the level of $\sim 10$\% in amplitude:
\BE
\label{eq:transfm}
T^{2} (f_{\rm Ch}, k_{\rm d}) = (1 - f_{\rm Ch}) + \frac{f_{\rm Ch}}{(1 + k / k_{\rm d})^{0.7441}} \,, \quad
f_{\rm Ch} (r_{\rm Ch}) = 1 - \exp \left(- 1.382 \frac{r_{\rm Ch}^{0.5168}}{1 - r_{\rm Ch}^{0.9417}} \right) \,,
\EE
where $k_{\rm d}$ is the same as given in eq.\,(\ref{eq:transfc}).
We assume a functional form of $f_{\rm Ch}$ as $ f_{\rm Ch} ( r_{\rm Ch}) = 1  - \exp[ -a r_{\rm Ch}^b / (1- r_{\rm Ch}^c) ]$ by analogy to $f_{\rm warm}$ in \cite{Kamada:2016vsc}, and then vary and determine the three parameters ($a, b,$ and $c$) by minimizing the residual given in eq.\,(\ref{eq:res}) with the sum over the mixed long-lived CHAMP models ($r_{\rm Ch} < 1$) listed in table\,\ref{tab:modpara}.

The comparison between $T^{2}$ and $P_{\rm MDM}/P_{\rm CDM} |_{\rm L5, L10}$ is given in figure\,\ref{fig:transfm}.
Note that the MDM model with $(m_{\rm WDM} \, [{\rm keV}], r_{\rm warm}) = (1, 0.8)$ (blue), which shows the comparable transfer function to that in the CH(1,\,0.85) model, is disfavored at the 95\% confidence level.
Thus we expect that the CH(1,\,0.85) model is also excluded at the 95\% confidence level, which provides a consistency check for our analysis below.

\BF
\IG[width=0.5\linewidth]{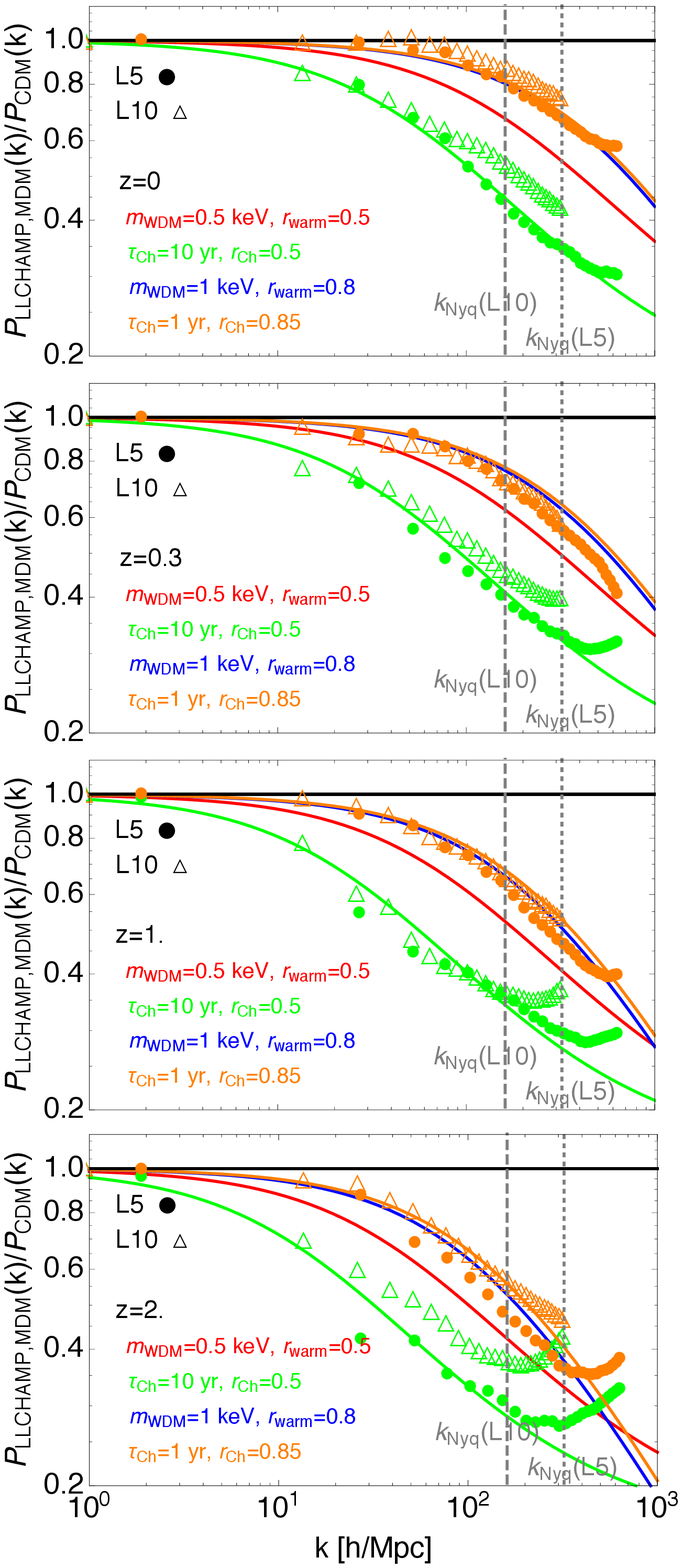}
\caption{The same figure as figure\,\ref{fig:transfc}, but in the mixed long-lived CHAMP models of CH(10,\,0.5) (green) and CH(1,\,0.85) (orange). 
See eq.\,(\ref{eq:transfm}) for $T^{2}$. 
This gives a reasonable fit up to 20\% in amplitude at low redshifts ($z < 2$).
The fitting formulas in the MDM models with $(m_{\rm WDM} \, [{\rm keV}], r_{\rm warm}) = (0.5, 0.5)$ (red) and $(1, 0.8)$ (blue) (see \cite{Kamada:2016vsc}) are also shown for reference.
}
\label{fig:transfm}
\EF

We construct the fitting formula of the non-linear matter power spectra in mixed long-lived CHAMP models by multiplying $T^{2}$ (eq.\,(\ref{eq:transfm})) and \texttt{halofit}, which has been developed for the standard CDM model with the same cosmological parameters as employed in this paper.
The explicit expression of \texttt{halofit} can be found in \cite{Takahashi:2012em, Inoue:2014jka} and thus is not repeated here.
The agreement between \texttt{halofit} and simulations with our setups is checked and confirmed in \cite{Kamada:2016vsc}.
We compare the fitting formula of the non-linear matter power spectra and the simulations in figures\,\ref{fig:nonlinHc} and \ref{fig:nonlinHm}.
Here we further check the fitting formula by using the LH10 simulations.
The agreement between HL10 and L10 sumulations is within at worst 10\% level in the non-linear matter powers, while L5 is 10\% smaller than these two.
A net error of the fitting formula can be estimated at 20\% at most in amplitude at $z < 2$.

\BF
\IG[width=0.5\linewidth]{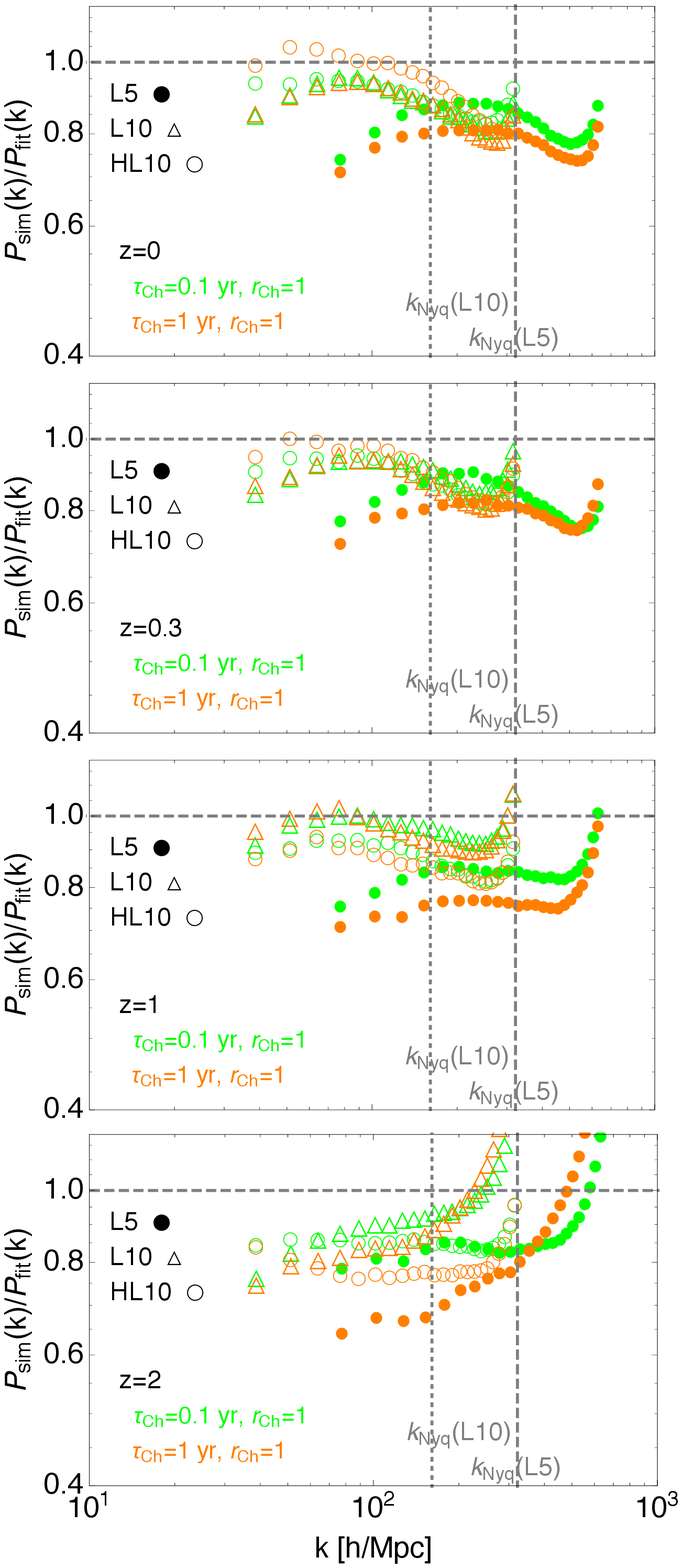}
\caption{$P_{\rm MDM}|_{\rm L5, L10, HL10}/P_{\rm MDM}|_{\texttt{halofit}\, {\rm multiplied \, by}\, T^{2}}$ at $z=0,0.3,1,$ and $2$. 
We show the CH(0.1,\,1) (green) and CH(1,\,1) (orange) models.
}
\label{fig:nonlinHc}
\EF

\BF
\IG[width=0.5\linewidth]{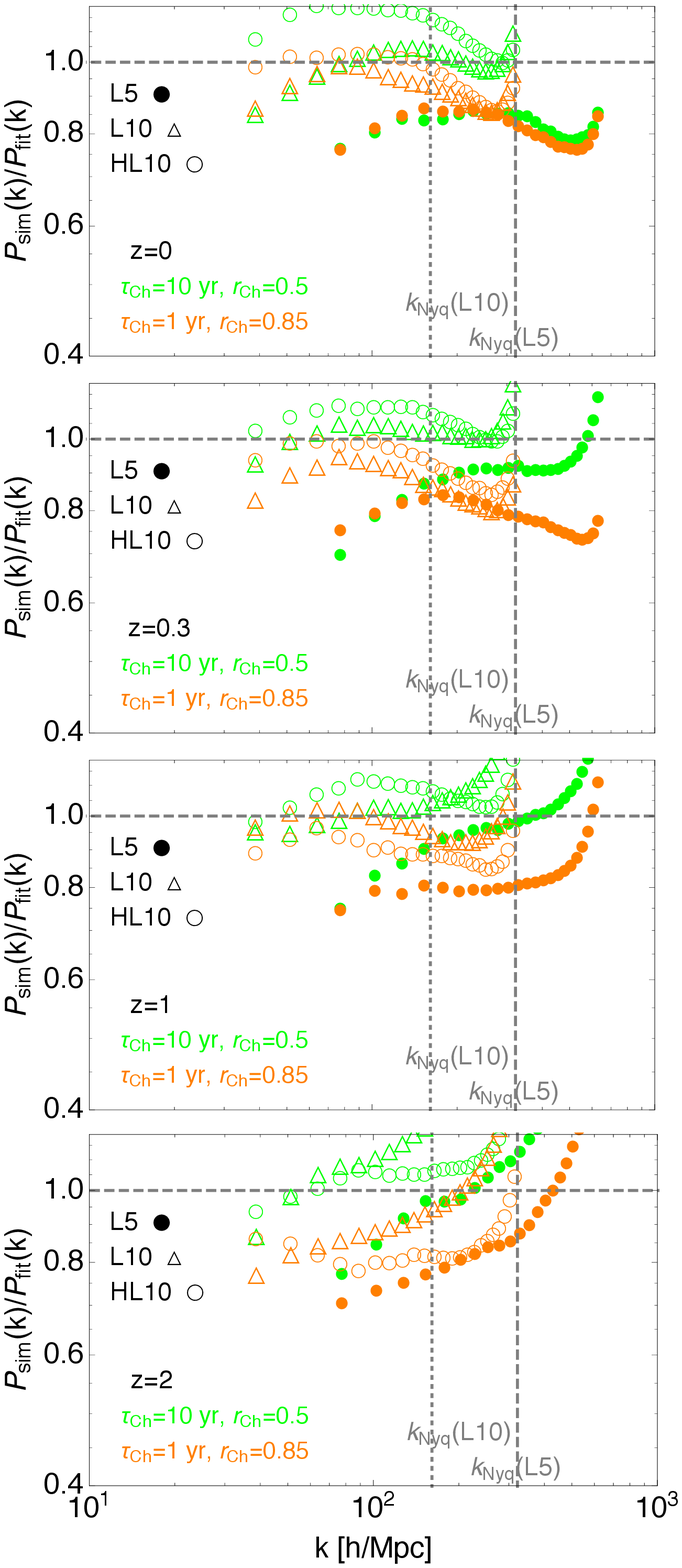}
\caption{$P_{\rm MDM}|_{\rm L5, L10, HL10}/P_{\rm MDM}|_{\texttt{halofit}\, {\rm multiplied \, by}\, T^{2}}$ at $z=0,0.3,1,$ and $2$. 
We show the CH(1,\,0.85) (green) and CH(10,\,0.5) (orange) models.
}
\label{fig:nonlinHm}
\EF

\section{Lens analysis and result}
\label{sec:lenanares}
The anomalous quadruple lens samples and the analyzing method are the same as in \cite{Inoue:2014jka, Kamada:2016vsc}.
The samples are B1422$+$231, B0128$+$437, MG0414$+$0534, and B0712$+$472.
The primary lens galaxy halo is modeled by a singular isothermal ellipsoid. 
The parameters in the model are determined by fitting the positions of lensed images and the centroid of the primary lensing galaxy.
The derived flux ratio does not fit the observed data at the 95\% confidence level or more.

To quantify the probability of reproducing the observed flux ratios in a given model, we use the method derived in \cite{Inoue:2012px, Takahashi:2013sna}.
Below we briefly describe the methodology of our analysis. 
For details, we refer the readers to \cite{Inoue:2012px, Takahashi:2013sna}.
The magnification perturbation can be characterized by an estimator of
\BE
\label{eta_def}
\eta \equiv \biggl[\frac{1}{2 N_{\rm pair}} \sum_{ i \neq j} \left[ \delta^\mu_{i} (minimum) - \delta^\mu_{ j} (saddle) \right]^2 \biggr]^{1/2} \,,
\EE
in which the sum is taken over $N_{\rm pair}$ pairs of images of a point source with different parities and $\delta^\mu_i = \delta \mu_i/\mu_i$ with $\mu_{i}$ and $\delta \mu_{i}$ being the magnification and its perturbation of the image {\it $i$}, which is evaluated at {\it minimum} and {\it saddle} points in the arrival time surface.

To obtain constraints on the CHAMP model parameters ($\tau_{\rm Ch}$ and $r_{\rm Ch}$), we calculate the $p$-value with the following formula:
\BE
p (\tau_{\rm Ch}, r_{\rm Ch}) = \left( \int_{{\hat V}} \prod_{a} d \eta_{a} P (\eta_{a}; \langle \eta_{a}^{2} \rangle^{1/2}, \delta {\hat \eta}_{a})\right) {\bigg /} \left( \int \prod_{a} d \eta_{a} P (\eta_{a}; \langle \eta_{a}^{2} \rangle^{1/2}, \delta {\hat \eta}_{a}) \right) \,,
\EE
where $P (\eta_{a}; \langle \eta_{a}^{2} \rangle^{1/2}, \delta {\hat \eta}_{a}) $ is the PDF for the anomalous lens system $a$ with $\eta_a$ and $\langle \eta_{a}^{2} \rangle$ being respectively the estimator for the system $a$ and the ensemble average of its second moment.
$\delta {\hat \eta}$ is the observational error which is taken into account by replacing $\langle \eta^{2} \rangle^{1/2}$ as $\langle \eta^{2} \rangle^{1/2} \to (\langle \eta^{2} \rangle + \delta {\hat \eta}^{2})^{1/2}$ in the PDF.
We define the integration domain (${\hat V}$) by
\BE
\prod_{a} P (\eta_{a} \in {\hat V}; \langle \eta_{a}^{2} \rangle^{1/2}, \delta {\hat \eta}_{a}) < \prod_{a} P ({\hat \eta}_{a}; \langle \eta_{a}^{2} \rangle^{1/2}, \delta {\hat \eta}_{a}) \,,
\EE
with the observed value of the estimator being ${\hat \eta}$.
For the PDF of $\eta$, we assume a lognormal form as follows:
\BE
\label{eq:PDF}
P(\eta; \langle \eta^{2} \rangle^{1/2}) \propto \exp\left[- \left\{ \ln(1+\eta/\eta_{0} (\langle \eta^{2} \rangle^{1/2})) -\ln(\mu) \right\} / (2\sigma^2) \right] / (\eta + \eta_{0}) \,.
\EE
Here three parameters $\eta_0, \mu$, and $\sigma$ are taken to be 
\BE
\label{eq:PDF_param}
 \eta_{0} (\langle \eta^{2} \rangle^{1/2}) = 0.228 \langle \eta^{2} \rangle^{1/2}, \quad \mu=4.10, \quad \sigma^2 = 0.279 \,,
\EE
which have been calibrated to the PDF obtained by ray-tracing simulations in the standard CDM model\,\cite{Takahashi:2013sna}.
We adopt the values given in eq.\,(\ref{eq:PDF_param}), which is considered to be a reasonable assumption in WDM models\,\cite{Inoue:2014jka}, also in long-lived CHAMP models.

Here we adopt the weak lens approximation, where $|\delta_i^{\mu}| \ll 1$.
For the case with three images A, B, and C where A and B are minimum points and C is a saddle one, the second moment of the estimator can be calculated as
\BE
\label{eq:estimator-anal}
\langle \eta^2 \rangle = \frac{1}{4}\biggl[(I_A+I_B)-2I_{AB}(\theta_{AB})+(I_B+I_C) - 2I_{BC}(\theta_{BC}) \biggr] \,,
\EE
where $\theta_{ij}$ $(i,j = \tr{A,B,C})$ corresponds to the separation angles between the images $i$ and $j$.
Here $ I_{ij}(\theta_{ij})$ is defined by 
\BEA
\label{eq:Iij}
I_{ij}(\theta_{ij}) &=& 4 \mu_i \mu_j \biggl[(1-\kappa_i)(1-\kappa_j) \langle \delta \kappa (0) \delta \kappa(\theta_{ij}) \rangle 
+ \gamma_{1i}\gamma_{1j} \langle \delta \gamma_1 (0) \delta \gamma_1 (\theta_{ij}) \rangle +\gamma_{2i}\gamma_{2j} \langle \delta \gamma_2 (0) \delta \gamma_2(\theta_{ij}) \rangle \N \\
&& + (1-\kappa_i)\gamma_{1j} \langle \delta \kappa_i(0) \delta \gamma_{1j} (\theta_{ij}) \rangle
+ (1-\kappa_j)\gamma_{1i} \langle \delta \kappa_j(0) \delta \gamma_{1i}(\theta_{ij}) \rangle
\biggr],
\EEA
where $\kappa_i$ is the convergence, $\gamma_{1i}$ and $\gamma_{2i}$ are the components of the shear, and $\delta \kappa_i, \delta \gamma_{1i}$, and $\delta \gamma_{2i}$ are their perturbations of the image $i$. 
Here we take the coordinates such that $\langle \delta \kappa \delta \gamma_2 \rangle$ and $\langle \delta \gamma_1 \delta \gamma_2 \rangle$ vanish by setting the separation angle between a pair of lensed images perpendicular to the $+$ mode.
$I_{i}$ $(i=\tr{A,B,C})$ is defined by using eq.\,(\ref{eq:Iij}) as $I_{i}= I_{ii}(0)$.

The non-linear matter power spectrum ($P_\delta (k,r)$) at the comoving distance of $r$ is related to the auto-correlation function of $\delta \kappa$ as follows:
\BE
\label{eq:ck}
\langle \delta \kappa (0) \delta \kappa (\theta) \rangle = \frac{9 H_0^4 \Omega_{m,0}^2}{4 c^4} \int_0^{r_S} dr  r^2 \biggl(\frac{r-r_{\rm S}}{r_{\rm S}} \biggr)^2 [1+z(r)]^2 \int_{k_{\rm {lens}}}^{k_{\tr{max}}}\frac{dk}{2 \pi} k W_{\textrm{CS}}^2(k;k_{\textrm{cut}})  P_{\delta}(k,r) J_0(g(r) k\theta) \,,
\EE
where $z(r)$ is the redshift to the comoving distance and $J_{n}$ is the $n$th order Bessel function. 
$g(r)$ is given by 
\BE
g(r)= \left\{ 
\begin{array}{ll}
r & \mbox{for $r<r_{\rm L}$} \\
{r_L(r_S-r)}/{(r_S-r_L)} & \mbox{for $r\ge r_{\rm L}$}
\end{array}
\,, \right.
\label{eq:g}
\EE
which describes the photon trajectory in a primary lens, where $r_L$ and $r_S$ represent respectively the comoving distances to the lens galaxy and the source.
In the $k$-integral of eq.\,(\ref{eq:ck}), the lower bound is given by $k_{\tr{lens}} = \pi/ (2 r_{\rm L} b)$ with $b$ being the mean angular separation between a lensed image and a lens centre.
The upper bound is chosen to be $k_{\tr{max}} = \mathcal{O}(10^{3 \text{--}5}) \, h/$Mpc, which corresponds to the scale above which perturbations become negligible due to the finite source size ($\sim \mathcal{O}(1)$\,pc).
$W_{\rm CS}$ is a filtering function which is called the {\it constant shift} (CS) filter\,\cite{Takahashi:2013sna} and mildly cuts the contribution from large angular-scale modes with $k < k_{\rm cut}$.
Other correlation functions $\langle \delta \gamma_1(0) \delta \gamma_1(\theta) \rangle$, $\langle \delta \gamma_2(0) \delta \gamma_2(\theta) \rangle$, and $\langle \delta \kappa(0) \delta \gamma_1(\theta) \rangle$ are also given by similar expressions whose explicit forms can be found in \cite{Inoue:2014jka} and thus are not repeated here.

We compute the non-linear matter power spectra (\texttt{halofit} multiplied by $T^{2}$ (eq.\,(\ref{eq:transfm}))) in 25 models including the long-lived CHAMP (total 15) models listed in table\,\ref{tab:modpara}.
We calculate $\langle \eta_{a}^{2} \rangle^{1/2} (\tau_{\rm Ch}, r_{\rm Ch})$ for each lens system $a$ from eq.\,(\ref{eq:estimator-anal}).
We interpolate the resultant $p$-values linearly in the $(\tau_{\rm Ch}, r_{\rm Ch})$ plane and show the result in figure\,\ref{fig:pllchamp}.
The pure long-lived CHAMP models with $\tau_{\rm Ch} > 0.96$\,yr fail in reproducing the observed flux ratios, $p(\tau_{\rm Ch}, r_{\rm Ch} = 1) < 0.05$.
The constraint on $\tau_{\rm Ch}$ becomes weaker for smaller $r_{\rm Ch}$.
This is because for a given $\tau_{\rm Ch}$, the suppression of matter power spectra is milder for smaller $r_{\rm Ch}$.
The CH(1,\,0.85) model is excluded as expected from its similar non-linear matter power spectrum to that in the MDM model with $(m_{\rm WDM} \, [{\rm keV}], r_{\rm warm}) = (1, 0.8)$ (see discussion in section\,\ref{subsec:nonlinpower}).
The mixed long-lived CHAMP models with the lifetime as long as $\tau_{\rm Ch} = 10$\,yr are compatible if $r_{\rm Ch} < 0.5$ at the 95\% confidence level.

\BF
\IG[width=0.75\linewidth]{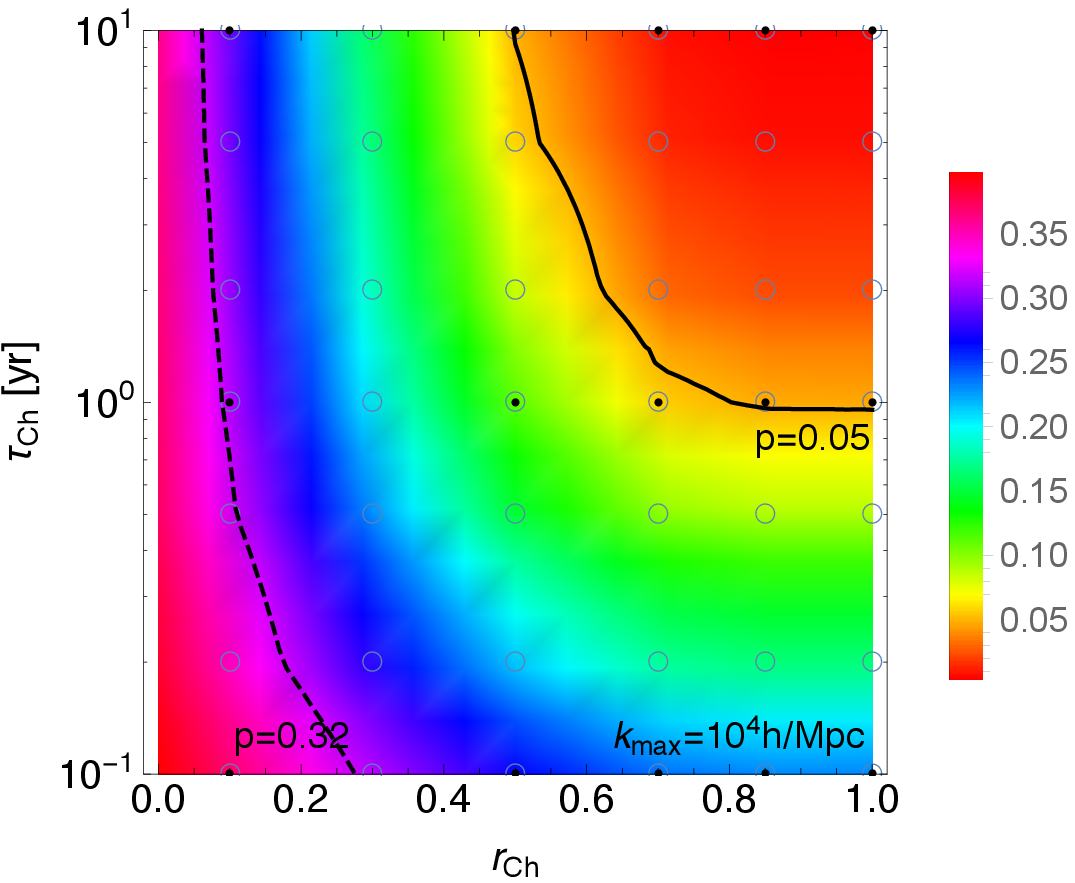}
\caption{$p$-value as a function of $\tau_{\rm Ch}$ and $r_{\rm Ch}$. 
We set a cut-off scale of $k_{\rm max}=10^4\,h/$Mpc. 
Circles denote the parameter sets where we evaluate $p$-values directly through the \texttt{halofit} multiplied by the fitting formula. 
These $p$-values are interpolated linearly to those at other parameter sets.
Dots in circles indicate that the models are listed in table\,\ref{tab:modpara}, in which $N$-body simulations are performed.
}
\label{fig:pllchamp}
\EF

\section{Conclusion}
\label{sec:concl}
We have examined the possibility of reproducing the observed flux ratios in four anomalous quadruple lens systems via weak lens effects of the line-of-sight matter distribution in long-lived CHAMP models.
In long-lived CHAMP models the density perturbations at subgalactic scales are suppressed as in WDM or MDM models, whose behavior of the magnification perturbations was studied in \cite{Inoue:2014jka, Kamada:2016vsc}.
Although our method for constraining CHAMP models is similar to that used for constraining MDM models, an independent analysis was required.
In the two models, the suppression mechanisms of the matter power spectra are different:
the acoustic damping in long-lived CHAMP models and the collisionless damping in MDM models.
In fact, we can map WDM models to pure long-lived CHAMP models as suggested in the previous literature, but the mapping of the model parameters between mixed long-lived CHAMP models and MDM models is non-trivial.
This point has been clearly seen in our calculated linear matter power spectra.
Meanwhile the mapping of the model parameters between pure long-lived CHAMP models and WDM models has been found valid in the non-linear matter power spectra, or in other words, even after the non-linear gravitational evolution.

Our analysis has shown that the pure long-lived CHAMP models with $\tau_{\rm Ch} < 0.96$\,yr are successful in reproducing the observed flux ratios at the 95\% confidence level or more.
A shorter CHAMP lifetime implies an earlier realization of the CDM model and thus a wider range matches of the matter power spectra between long-lived CHAMP models and the standard CDM model.
An upper bound on the CHAMP lifetime corresponds to a lower bound on the mass of the warm component in WDM models, where the CDM model is the heavy mass limit of the WDM model.
The mapping of the model parameters between them is compatible with the previous literature: $\tau_{\rm Ch} \simeq 1.5 \, {\rm yr} \, (1 \, {\rm keV} / m_{\rm WDM})^{8/3}$.

Once the mass density fraction of the long-lived CHAMP component is reduced and some CDM-like particle is introduced to account for the rest, the constraint on the CHAMP lifetime has become weaker.  
This is because the suppressions in linear matter power spectra become $r_{\rm Ch}^{2}$ times milder when compared to pure long-lived CHAMP models ($r_{\rm Ch} = 1$) with the same CHAMP lifetime.  
The derived constraints on the long-lived CHAMP mass density fraction is $r_{\rm Ch} < 0.5$ (95\% confidence level) for the CHAMP lifetime of $\tau_{\rm Ch} = 10$\,yr.

In order to obtain such a long lifetime to be sensitive to the current analyses in concrete models, e.g., in the MSSM, there are two possible scenarios of the stau NLSP decaying into a LSP dark matter particle: 1) into the gravitino LSP and 2) into the neutralino LSP with a small mass difference between the stau and the neutralino. 
In either scenario, the masses should be commonly degenerate between the CHAMP and the daughter particle to be $m_{\rm Ch} \sim m_{\rm daughter}$, which is a general condition to avoid the photodissociation of light elements\,\cite{Kawasaki:2004qu, Jedamzik:2006xz, Kawasaki:2008qe} and the CMB distortion\,\cite{Hu:1993gc,Chluba:2011hw}, independently of particle physics models. 
Regarding the CHAMP abundance, when we adopt a non-thermal production of CHAMPs, we do not have to stick to the standard thermal relic abundance of CHAMPs. 
However, we have to be careful for the condition that the mass density of daughter particles produced by decays of CHAMPs should not exceed the observed DM mass density, i.e., $\Omega_{\rm daughter} \sim \Omega_{\rm Ch} = r_{\rm Ch} \Omega_{c,0}$.
Meanwhile, to evade the constraint from the catalyzed effect on an overproduction of the $^6$Li abundance\,\cite{Pospelov:2006sc, Steffen:2006wx, Hamaguchi:2007mp, Bird:2007ge, Kawasaki:2007xb, Kawasaki:2008qe}, we have a strict big-bang nucleosynthesis bound on the ratio of the number density of CHAMPs to the entropy density ($s$): $n_{\rm Ch}/s \lesssim 10^{-15}$ for $\tau_{\rm Ch} \gtrsim 10^3$\,s. 
In summary, we have a lower bound on the mass of the CHAMP, $m_{\rm Ch} \gtrsim 4 \times 10^2 r_{\rm Ch}$\,TeV, which can be realized in concrete models with $r_{\rm Ch} < 1$, e.g., in the MSSM.

Although, we have used only four samples of quadruple lens systems with flux-ratio anomalies, the number of such systems will surge in near future. 
For instance, the recently found quadruple lens systems of submillimeter galaxies\,\cite{Negrello:2010qw} and Lyman-alpha emitting galaxies\,\cite{2016ApJ...833..264S} are expected to contain many systems with flux-ratio anomalies. 
Using such new samples, we would have a more stringent constraint on the long-lived CHAMP model parameters.

\acknowledgments
This work was partially supported by IBS under the project code IBS-R018-D1 (AK), JSPS KAKENHI Grant Numbers 26247042 (KK) and 15K05084 (TT), and MEXT KAKENHI Grant Numbers 15H05889, 16H00877 (KK), and 15H05888 (TT).
Numerical computations were carried out on Cray XC30 at Center for Computational Astrophysics, National Astronomical Observatory of Japan and on SR16000 at YITP in Kyoto University.

\appendix









\bibliographystyle{JHEP}
\bibliography{LLCHAMP.bib}

\providecommand{\href}[2]{#2}\begingroup\raggedright\begin{thebibliography}{10}

\bibitem{Sigurdson:2003vy}
K.~Sigurdson and M.~Kamionkowski, \emph{{Charged - particle decay and
  suppression of small - scale power}},
  \href{http://dx.doi.org/10.1103/PhysRevLett.92.171302}{\emph{Phys. Rev.
  Lett.} {\bf 92} (2004) 171302},
  [\href{http://arxiv.org/abs/astro-ph/0311486}{{\tt astro-ph/0311486}}].

\bibitem{Profumo:2004qt}
S.~Profumo, K.~Sigurdson, P.~Ullio and M.~Kamionkowski, \emph{{A Running
  spectral index in supersymmetric dark-matter models with quasi-stable charged
  particles}}, \href{http://dx.doi.org/10.1103/PhysRevD.71.023518}{\emph{Phys.
  Rev.} {\bf D71} (2005) 023518},
  [\href{http://arxiv.org/abs/astro-ph/0410714}{{\tt astro-ph/0410714}}].

\bibitem{Kohri:2009mi}
K.~Kohri and T.~Takahashi, \emph{{Cosmology with Long-Lived Charged Massive
  Particles}},
  \href{http://dx.doi.org/10.1016/j.physletb.2009.11.051}{\emph{Phys. Lett.}
  {\bf B682} (2010) 337--341}, [\href{http://arxiv.org/abs/0909.4610}{{\tt
  0909.4610}}].

\bibitem{Kamada:2013sh}
A.~Kamada, N.~Yoshida, K.~Kohri and T.~Takahashi, \emph{{Structure of Dark
  Matter Halos in Warm Dark Matter models and in models with Long-Lived Charged
  Massive Particles}},
  \href{http://dx.doi.org/10.1088/1475-7516/2013/03/008}{\emph{JCAP} {\bf 1303}
  (2013) 008}, [\href{http://arxiv.org/abs/1301.2744}{{\tt 1301.2744}}].

\bibitem{Moore:1999nt}
B.~Moore, S.~Ghigna, F.~Governato, G.~Lake, T.~R. Quinn, J.~Stadel et~al.,
  \emph{{Dark matter substructure within galactic halos}},
  \href{http://dx.doi.org/10.1086/312287}{\emph{\apj} {\bf 524} (1999)
  L19--L22}, [\href{http://arxiv.org/abs/astro-ph/9907411}{{\tt
  astro-ph/9907411}}].

\bibitem{Kravtsov:2009gi}
A.~V. Kravtsov, \emph{{Dark matter substructure and dwarf galactic
  satellites}}, \href{http://dx.doi.org/10.1155/2010/281913}{\emph{Adv.
  Astron.} {\bf 2010} (2010) 281913},
  [\href{http://arxiv.org/abs/0906.3295}{{\tt 0906.3295}}].

\bibitem{Pospelov:2006sc}
M.~Pospelov, \emph{{Particle physics catalysis of thermal Big Bang
  Nucleosynthesis}},
  \href{http://dx.doi.org/10.1103/PhysRevLett.98.231301}{\emph{Phys. Rev.
  Lett.} {\bf 98} (2007) 231301},
  [\href{http://arxiv.org/abs/hep-ph/0605215}{{\tt hep-ph/0605215}}].

\bibitem{Steffen:2006wx}
F.~D. Steffen, \emph{{Constraints on Gravitino Dark Matter Scenarios with
  Long-Lived Charged Sleptons}},
  \href{http://dx.doi.org/10.1063/1.2735255}{\emph{AIP Conf. Proc.} {\bf 903}
  (2007) 595--598}, [\href{http://arxiv.org/abs/hep-ph/0611027}{{\tt
  hep-ph/0611027}}].

\bibitem{Hamaguchi:2007mp}
K.~Hamaguchi, T.~Hatsuda, M.~Kamimura, Y.~Kino and T.~T. Yanagida,
  \emph{{Stau-catalyzed Li-6 Production in Big-Bang Nucleosynthesis}},
  \href{http://dx.doi.org/10.1016/j.physletb.2007.05.030}{\emph{Phys. Lett.}
  {\bf B650} (2007) 268--274}, [\href{http://arxiv.org/abs/hep-ph/0702274}{{\tt
  hep-ph/0702274}}].

\bibitem{Bird:2007ge}
C.~Bird, K.~Koopmans and M.~Pospelov, \emph{{Primordial Lithium Abundance in
  Catalyzed Big Bang Nucleosynthesis}},
  \href{http://dx.doi.org/10.1103/PhysRevD.78.083010}{\emph{Phys. Rev.} {\bf
  D78} (2008) 083010}, [\href{http://arxiv.org/abs/hep-ph/0703096}{{\tt
  hep-ph/0703096}}].

\bibitem{Kawasaki:2007xb}
M.~Kawasaki, K.~Kohri and T.~Moroi, \emph{{Big-Bang Nucleosynthesis with
  Long-Lived Charged Slepton}},
  \href{http://dx.doi.org/10.1016/j.physletb.2007.03.063}{\emph{Phys. Lett.}
  {\bf B649} (2007) 436--439}, [\href{http://arxiv.org/abs/hep-ph/0703122}{{\tt
  hep-ph/0703122}}].

\bibitem{Kawasaki:2008qe}
M.~Kawasaki, K.~Kohri, T.~Moroi and A.~Yotsuyanagi, \emph{{Big-Bang
  Nucleosynthesis and Gravitino}},
  \href{http://dx.doi.org/10.1103/PhysRevD.78.065011}{\emph{Phys. Rev.} {\bf
  D78} (2008) 065011}, [\href{http://arxiv.org/abs/0804.3745}{{\tt
  0804.3745}}].

\bibitem{Kamada:2016qjo}
A.~Kamada, K.~Kohri, T.~Takahashi and N.~Yoshida, \emph{{Effects of
  electrically charged dark matter on cosmic microwave background
  anisotropies}},  \href{http://arxiv.org/abs/1604.07926}{{\tt 1604.07926}}.

\bibitem{Kadota:2016tqq}
K.~Kadota, T.~Sekiguchi and H.~Tashiro, \emph{{A new constraint on millicharged
  dark matter from galaxy clusters}},
  \href{http://arxiv.org/abs/1602.04009}{{\tt 1602.04009}}.

\bibitem{Feng:2003xh}
J.~L. Feng, A.~Rajaraman and F.~Takayama, \emph{{Superweakly interacting
  massive particles}},
  \href{http://dx.doi.org/10.1103/PhysRevLett.91.011302}{\emph{Phys. Rev.
  Lett.} {\bf 91} (2003) 011302},
  [\href{http://arxiv.org/abs/hep-ph/0302215}{{\tt hep-ph/0302215}}].

\bibitem{Feng:2003uy}
J.~L. Feng, A.~Rajaraman and F.~Takayama, \emph{{SuperWIMP dark matter signals
  from the early universe}},
  \href{http://dx.doi.org/10.1103/PhysRevD.68.063504}{\emph{Phys. Rev.} {\bf
  D68} (2003) 063504}, [\href{http://arxiv.org/abs/hep-ph/0306024}{{\tt
  hep-ph/0306024}}].

\bibitem{Feng:2004zu}
J.~L. Feng, S.-f. Su and F.~Takayama, \emph{{SuperWIMP gravitino dark matter
  from slepton and sneutrino decays}},
  \href{http://dx.doi.org/10.1103/PhysRevD.70.063514}{\emph{Phys. Rev.} {\bf
  D70} (2004) 063514}, [\href{http://arxiv.org/abs/hep-ph/0404198}{{\tt
  hep-ph/0404198}}].

\bibitem{Feng:2004mt}
J.~L. Feng, S.~Su and F.~Takayama, \emph{{Supergravity with a gravitino LSP}},
  \href{http://dx.doi.org/10.1103/PhysRevD.70.075019}{\emph{Phys. Rev.} {\bf
  D70} (2004) 075019}, [\href{http://arxiv.org/abs/hep-ph/0404231}{{\tt
  hep-ph/0404231}}].

\bibitem{Dine:1994vc}
M.~Dine, A.~E. Nelson and Y.~Shirman, \emph{{Low-energy dynamical supersymmetry
  breaking simplified}},
  \href{http://dx.doi.org/10.1103/PhysRevD.51.1362}{\emph{Phys. Rev.} {\bf D51}
  (1995) 1362--1370}, [\href{http://arxiv.org/abs/hep-ph/9408384}{{\tt
  hep-ph/9408384}}].

\bibitem{Dine:1995ag}
M.~Dine, A.~E. Nelson, Y.~Nir and Y.~Shirman, \emph{{New tools for low-energy
  dynamical supersymmetry breaking}},
  \href{http://dx.doi.org/10.1103/PhysRevD.53.2658}{\emph{Phys. Rev.} {\bf D53}
  (1996) 2658--2669}, [\href{http://arxiv.org/abs/hep-ph/9507378}{{\tt
  hep-ph/9507378}}].

\bibitem{Jittoh:2007fr}
T.~Jittoh, K.~Kohri, M.~Koike, J.~Sato, T.~Shimomura and M.~Yamanaka,
  \emph{{Possible solution to the Li-7 problem by the long lived stau}},
  \href{http://dx.doi.org/10.1103/PhysRevD.76.125023}{\emph{Phys. Rev.} {\bf
  D76} (2007) 125023}, [\href{http://arxiv.org/abs/0704.2914}{{\tt
  0704.2914}}].

\bibitem{Kawasaki:2004qu}
M.~Kawasaki, K.~Kohri and T.~Moroi, \emph{{Big-Bang nucleosynthesis and
  hadronic decay of long-lived massive particles}},
  \href{http://dx.doi.org/10.1103/PhysRevD.71.083502}{\emph{Phys. Rev.} {\bf
  D71} (2005) 083502}, [\href{http://arxiv.org/abs/astro-ph/0408426}{{\tt
  astro-ph/0408426}}].

\bibitem{Jedamzik:2006xz}
K.~Jedamzik, \emph{{Big bang nucleosynthesis constraints on hadronically and
  electromagnetically decaying relic neutral particles}},
  \href{http://dx.doi.org/10.1103/PhysRevD.74.103509}{\emph{Phys. Rev.} {\bf
  D74} (2006) 103509}, [\href{http://arxiv.org/abs/hep-ph/0604251}{{\tt
  hep-ph/0604251}}].

\bibitem{Inoue:2014jka}
K.~T. Inoue, R.~Takahashi, T.~Takahashi and T.~Ishiyama, \emph{{Constraints on
  warm dark matter from weak lensing in anomalous quadruple lenses}},
  \href{http://dx.doi.org/10.1093/mnras/stv194}{\emph{Mon. Not. Roy. Astron.
  Soc.} {\bf 448} (2015) 2704--2716},
  [\href{http://arxiv.org/abs/1409.1326}{{\tt 1409.1326}}].

\bibitem{Kamada:2016vsc}
A.~Kamada, K.~T. Inoue and T.~Takahashi, \emph{{Constraints on mixed dark
  matter from anomalous strong lens systems}},
  \href{http://dx.doi.org/10.1103/PhysRevD.94.023522}{\emph{Phys. Rev.} {\bf
  D94} (2016) 023522}, [\href{http://arxiv.org/abs/1604.01489}{{\tt
  1604.01489}}].

\bibitem{Mao:1997ek}
S.-d. Mao and P.~Schneider, \emph{{Evidence for substructure in lens
  galaxies?}},
  \href{http://dx.doi.org/10.1046/j.1365-8711.1998.01319.x}{\emph{Mon. Not.
  Roy. Astron. Soc.} {\bf 295} (1998) 587--594},
  [\href{http://arxiv.org/abs/astro-ph/9707187}{{\tt astro-ph/9707187}}].

\bibitem{Chiba:2001wk}
M.~Chiba, \emph{{Probing dark matter substructure in lens galaxies}},
  \href{http://dx.doi.org/10.1086/324493}{\emph{Astrophys. J.} {\bf 565} (2002)
  17}, [\href{http://arxiv.org/abs/astro-ph/0109499}{{\tt astro-ph/0109499}}].

\bibitem{Dalal:2001fq}
N.~Dalal and C.~S. Kochanek, \emph{{Direct detection of CDM substructure}},
  \href{http://dx.doi.org/10.1086/340303}{\emph{Astrophys. J.} {\bf 572} (2002)
  25--33}, [\href{http://arxiv.org/abs/astro-ph/0111456}{{\tt
  astro-ph/0111456}}].

\bibitem{Metcalf:2001es}
R.~B. Metcalf and H.~Zhao, \emph{{Flux ratios as a probe of dark substructures
  in quadruple-image gravitational lenses}},
  \href{http://dx.doi.org/10.1086/339798}{\emph{Astrophys. J.} {\bf 567} (2002)
  L5}, [\href{http://arxiv.org/abs/astro-ph/0111427}{{\tt astro-ph/0111427}}].

\bibitem{Kochanek:2003zc}
C.~S. Kochanek and N.~Dalal, \emph{{Tests for substructure in gravitational
  lenses}}, \href{http://dx.doi.org/10.1086/421436}{\emph{Astrophys. J.} {\bf
  610} (2004) 69--79}, [\href{http://arxiv.org/abs/astro-ph/0302036}{{\tt
  astro-ph/0302036}}].

\bibitem{Metcalf:2003sz}
R.~B. Metcalf, L.~A. Moustakas, A.~J. Bunker and I.~R. Parry,
  \emph{{Spectroscopic gravitational lensing and limits on the dark matter
  substructure in Q2237+0305}},
  \href{http://dx.doi.org/10.1086/383243}{\emph{Astrophys. J.} {\bf 607} (2004)
  43--59}, [\href{http://arxiv.org/abs/astro-ph/0309738}{{\tt
  astro-ph/0309738}}].

\bibitem{Chiba:2005et}
M.~Chiba, T.~Minezaki, N.~Kashikawa, H.~Kataza and K.~T. Inoue, \emph{{Subaru
  mid-infrared imaging of the quadruple lenses PG1115+080 and B1422+231: Limits
  on substructure lensing}},
  \href{http://dx.doi.org/10.1086/430403}{\emph{Astrophys. J.} {\bf 627} (2005)
  53--61}, [\href{http://arxiv.org/abs/astro-ph/0503487}{{\tt
  astro-ph/0503487}}].

\bibitem{Sugai:2007ic}
H.~Sugai, A.~Kawai, A.~Shimono, T.~Hattori, G.~Kosugi, N.~Kashikawa et~al.,
  \emph{{Integral Field Spectroscopy of the Quadruply Lensed Quasar 1RXS
  J1131-1231: New Light on Lens Substructures}},
  \href{http://dx.doi.org/10.1086/513731}{\emph{Astrophys. J.} {\bf 660} (2007)
  1016--1022}, [\href{http://arxiv.org/abs/astro-ph/0702392}{{\tt
  astro-ph/0702392}}].

\bibitem{More:2008qm}
A.~More, J.~P. McKean, S.~More, R.~W. Porcas, L.~V.~E. Koopmans and M.~A.
  Garrett, \emph{{The role of luminous substructure in the gravitational lens
  system MG 2016+112}},
  \href{http://dx.doi.org/10.1111/j.1365-2966.2008.14342.x}{\emph{Mon. Not.
  Roy. Astron. Soc.} {\bf 394} (2009) 174},
  [\href{http://arxiv.org/abs/0810.5341}{{\tt 0810.5341}}].

\bibitem{Minezaki:2009ek}
T.~Minezaki, M.~Chiba, N.~Kashikawa, K.~T. Inoue and H.~Kataza, \emph{{Subaru
  Mid-infrared Imaging of the Quadruple Lenses. II. Unveiling Lens Structure of
  MG0414+0534 and Q2237+030}},
  \href{http://dx.doi.org/10.1088/0004-637X/697/1/610}{\emph{Astrophys. J.}
  {\bf 697} (2009) 610--618}, [\href{http://arxiv.org/abs/0903.2535}{{\tt
  0903.2535}}].

\bibitem{Xu:2009ch}
D.~D. Xu, S.~Mao, J.~Wang, V.~Springel, L.~Gao, S.~D.~M. White et~al.,
  \emph{{Effects of Dark Matter Substructures on Gravitational Lensing: Results
  from the Aquarius Simulations}},
  \href{http://dx.doi.org/10.1111/j.1365-2966.2009.15230.x}{\emph{Mon. Not.
  Roy. Astron. Soc.} {\bf 398} (2009) 1235},
  [\href{http://arxiv.org/abs/0903.4559}{{\tt 0903.4559}}].

\bibitem{Xu:2010gs}
D.~Xu, S.~Mao, A.~Cooper, J.~Wang, L.~Gao, C.~Frenk et~al., \emph{{Substructure
  Lensing: Effects of Galaxies, Globular Clusters \& Satellite Streams}},
  \href{http://dx.doi.org/10.1111/j.1365-2966.2010.17235.x}{\emph{Mon. Not.
  Roy. Astron. Soc.} {\bf 408} (2010) 1721},
  [\href{http://arxiv.org/abs/1004.3094}{{\tt 1004.3094}}].

\bibitem{Inoue:2017zxb}
K.~T. Inoue, S.~Matsushita, T.~Minezaki and M.~Chiba, \emph{{Evidence for a
  Dusty Dark Dwarf Galaxy in the Quadruple Lens MG0414+0534}},
  \href{http://dx.doi.org/10.3847/2041-8213/835/2/L23}{\emph{Astrophys. J.}
  {\bf 835} (2017) L23}, [\href{http://arxiv.org/abs/1701.05283}{{\tt
  1701.05283}}].

\bibitem{Metcalf:2004eh}
R.~B. Metcalf, \emph{{The Importance of intergalactic structure to
  gravitationally lensed quasars}},
  \href{http://dx.doi.org/10.1086/431574}{\emph{Astrophys. J.} {\bf 629} (2005)
  673--679}, [\href{http://arxiv.org/abs/astro-ph/0412538}{{\tt
  astro-ph/0412538}}].

\bibitem{Xu:2011ru}
D.~D. Xu, S.~Mao, A.~Cooper, L.~Gao, C.~Frenk, R.~Angulo et~al., \emph{{On the
  Effects of Line-of-Sight Structures on Lensing Flux-ratio Anomalies in a LCDM
  Universe}},
  \href{http://dx.doi.org/10.1111/j.1365-2966.2012.20484.x}{\emph{Mon. Not.
  Roy. Astron. Soc.} {\bf 421} (2012) 2553},
  [\href{http://arxiv.org/abs/1110.1185}{{\tt 1110.1185}}].

\bibitem{Inoue:2012px}
K.~T. Inoue and R.~Takahashi, \emph{{Weak Lensing by Line-of-sight Halos as the
  Origin of Flux-ratio Anomalies in Quadruply Lensed QSOs}},
  \href{http://dx.doi.org/10.1111/j.1365-2966.2012.21915.x}{\emph{Mon. Not.
  Roy. Astron. Soc.} {\bf 426} (2012) 2978--2993},
  [\href{http://arxiv.org/abs/1207.2139}{{\tt 1207.2139}}].

\bibitem{Takahashi:2013sna}
R.~Takahashi and K.~T. Inoue, \emph{{Weak lensing by intergalactic
  ministructures in quadruple lens systems: simulation and detection}},
  \href{http://dx.doi.org/10.1093/mnras/stu328}{\emph{Mon. Not. Roy. Astron.
  Soc.} {\bf 440} (2014) 870--888}, [\href{http://arxiv.org/abs/1308.4855}{{\tt
  1308.4855}}].

\bibitem{Maccio:2005bj}
A.~V. Maccio and M.~Miranda, \emph{{The effect of low mass substructures on the
  cusp lensing relation}},
  \href{http://dx.doi.org/10.1111/j.1365-2966.2006.10154.x}{\emph{Mon. Not.
  Roy. Astron. Soc.} {\bf 368} (2006) 599--608},
  [\href{http://arxiv.org/abs/astro-ph/0509598}{{\tt astro-ph/0509598}}].

\bibitem{Amara:2004dr}
A.~Amara, R.~B. Metcalf, T.~J. Cox and O.~J. P., \emph{{Simulations of strong
  gravitational lensing with substructure}},
  \href{http://dx.doi.org/10.1111/j.1365-2966.2006.10053.x}{\emph{Mon. Not.
  Roy. Astron. Soc.} {\bf 367} (2006) 1367--1378},
  [\href{http://arxiv.org/abs/astro-ph/0411587}{{\tt astro-ph/0411587}}].

\bibitem{Chen:2008vt}
J.~Chen, \emph{{Parity Dependence in Strong Lens Systems as a Probe of Dark
  Matter Substructure}},
  \href{http://dx.doi.org/10.1051/0004-6361/200811134}{\emph{Astron.
  Astrophys.} {\bf 498} (2009) 49}, [\href{http://arxiv.org/abs/0810.2036}{{\tt
  0810.2036}}].

\bibitem{Chen:2011wc}
J.~Chen, S.~M. Koushiappas and A.~R. Zentner, \emph{{The Effects of
  Halo-to-Halo Variation on Substructure Lensing}},
  \href{http://dx.doi.org/10.1088/0004-637X/741/2/117}{\emph{Astrophys. J.}
  {\bf 741} (2011) 117}, [\href{http://arxiv.org/abs/1101.2916}{{\tt
  1101.2916}}].

\bibitem{Inoue:2014mla}
K.~T. Inoue, \emph{{Weak Lensing by Minifilament or Minivoid as the Origin of
  Flux-ratio Anomalies in Lensed Quasar MG0414+0534}},
  \href{http://dx.doi.org/10.1093/mnras/stu2507}{\emph{Mon. Not. Roy. Astron.
  Soc.} {\bf 447} (2015) 1452--1459},
  [\href{http://arxiv.org/abs/1410.1033}{{\tt 1410.1033}}].

\bibitem{Inoue:2016mqz}
K.~T. Inoue, \emph{{On the origin of the flux ratio anomaly in quadruple lens
  systems}}, \href{http://dx.doi.org/10.1093/mnras/stw1270}{\emph{Mon. Not.
  Roy. Astron. Soc.} {\bf 461} (2016) 164--175},
  [\href{http://arxiv.org/abs/1601.04414}{{\tt 1601.04414}}].

\bibitem{Inoue:2015lma}
K.~T. Inoue, T.~Minezaki, S.~Matsushita and M.~Chiba, \emph{{ALMA Imprint of
  Intergalactic Dark Structures in the Gravitational Lens SDP.81}},
  \href{http://dx.doi.org/10.1093/mnras/stw168}{\emph{Mon. Not. Roy. Astron.
  Soc.} {\bf 457} (2016) 2936--2950},
  [\href{http://arxiv.org/abs/1510.00150}{{\tt 1510.00150}}].

\bibitem{Miranda:2007rb}
M.~Miranda and A.~V. Maccio, \emph{{Constraining Warm Dark Matter using QSO
  gravitational lensing}},
  \href{http://dx.doi.org/10.1111/j.1365-2966.2007.12440.x}{\emph{Mon. Not.
  Roy. Astron. Soc.} {\bf 382} (2007) 1225},
  [\href{http://arxiv.org/abs/0706.0896}{{\tt 0706.0896}}].

\bibitem{Ade:2013zuv}
{\scshape Planck} collaboration, P.~A.~R. Ade et~al., \emph{{Planck 2013
  results. XVI. Cosmological parameters}},
  \href{http://dx.doi.org/10.1051/0004-6361/201321591}{\emph{Astron.
  Astrophys.} {\bf 571} (2014) A16},
  [\href{http://arxiv.org/abs/1303.5076}{{\tt 1303.5076}}].

\bibitem{Lewis:1999bs}
A.~Lewis, A.~Challinor and A.~Lasenby, \emph{{Efficient computation of CMB
  anisotropies in closed FRW models}},
  \href{http://dx.doi.org/10.1086/309179}{\emph{\apj} {\bf 538} (2000)
  473--476}, [\href{http://arxiv.org/abs/astro-ph/9911177}{{\tt
  astro-ph/9911177}}].

\bibitem{Hu:1993gc}
W.~Hu and J.~Silk, \emph{{Thermalization constraints and spectral distortions
  for massive unstable relic particles}},
  \href{http://dx.doi.org/10.1103/PhysRevLett.70.2661}{\emph{Phys. Rev. Lett.}
  {\bf 70} (1993) 2661--2664}.

\bibitem{Chluba:2011hw}
J.~Chluba and R.~A. Sunyaev, \emph{{The evolution of CMB spectral distortions
  in the early Universe}},
  \href{http://dx.doi.org/10.1111/j.1365-2966.2011.19786.x}{\emph{Mon. Not.
  Roy. Astron. Soc.} {\bf 419} (2012) 1294--1314},
  [\href{http://arxiv.org/abs/1109.6552}{{\tt 1109.6552}}].

\bibitem{Springel:2005mi}
V.~Springel, \emph{{The Cosmological simulation code GADGET-2}},
  \href{http://dx.doi.org/10.1111/j.1365-2966.2005.09655.x}{\emph{\mnras} {\bf
  364} (2005) 1105--1134}, [\href{http://arxiv.org/abs/astro-ph/0505010}{{\tt
  astro-ph/0505010}}].

\bibitem{Smith:2002dz}
{\scshape VIRGO Consortium} collaboration, R.~E. Smith, J.~A. Peacock,
  A.~Jenkins, S.~D.~M. White, C.~S. Frenk, F.~R. Pearce et~al., \emph{{Stable
  clustering, the halo model and nonlinear cosmological power spectra}},
  \href{http://dx.doi.org/10.1046/j.1365-8711.2003.06503.x}{\emph{Mon. Not.
  Roy. Astron. Soc.} {\bf 341} (2003) 1311},
  [\href{http://arxiv.org/abs/astro-ph/0207664}{{\tt astro-ph/0207664}}].

\bibitem{Wang:2007he}
J.~Wang and S.~D.~M. White, \emph{{Discreteness effects in simulations of
  Hot/Warm dark matter}},
  \href{http://dx.doi.org/10.1111/j.1365-2966.2007.12053.x}{\emph{Mon. Not.
  Roy. Astron. Soc.} {\bf 380} (2007) 93--103},
  [\href{http://arxiv.org/abs/astro-ph/0702575}{{\tt astro-ph/0702575}}].

\bibitem{Takahashi:2012em}
R.~Takahashi, M.~Sato, T.~Nishimichi, A.~Taruya and M.~Oguri, \emph{{Revising
  the Halofit Model for the Nonlinear Matter Power Spectrum}},
  \href{http://dx.doi.org/10.1088/0004-637X/761/2/152}{\emph{Astrophys. J.}
  {\bf 761} (2012) 152}, [\href{http://arxiv.org/abs/1208.2701}{{\tt
  1208.2701}}].

\bibitem{Negrello:2010qw}
M.~Negrello et~al., \emph{{The Detection of a Population of
  Submillimeter-Bright, Strongly-Lensed Galaxies}},
  \href{http://dx.doi.org/10.1126/science.1193420}{\emph{Science} {\bf 330}
  (2010) 800}, [\href{http://arxiv.org/abs/1011.1255}{{\tt 1011.1255}}].

\bibitem{2016ApJ...833..264S}
Y.~{Shu}, A.~S. {Bolton}, S.~{Mao}, C.~S. {Kochanek}, I.~{P{\'e}rez-Fournon},
  M.~{Oguri} et~al., \emph{{The BOSS Emission-line Lens Survey. IV. Smooth Lens
  Models for the BELLS GALLERY Sample}},
  \href{http://dx.doi.org/10.3847/1538-4357/833/2/264}{\emph{\apj} {\bf 833}
  (Dec., 2016) 264}, [\href{http://arxiv.org/abs/1608.08707}{{\tt
  1608.08707}}].

\end{thebibliography}\endgroup
 
\end{document}